\documentclass[endnotes]{article}
\usepackage{setspace}
\usepackage{endnotes}
\usepackage{amsmath}
\usepackage{graphicx}
\usepackage{color}
\usepackage[round]{natbib}

\begin{document}


\title{A Market for Unbiased Private Data:\\
\large{Paying Individuals According to their Privacy Attitudes}}
\date{}
\author{Christina Aperjis\\Social Computing Group\\HP Labs\\christina.aperjis@hp.com \and Bernardo A. Huberman\\Social Computing Group\\HP Labs\\bernardo.huberman@hp.com}

\maketitle





\begin{abstract}
Since there is, in principle, no reason why third parties should not pay individuals for the use of their data, we introduce a realistic market that would allow these payments to be made while taking into account the privacy attitude of the participants. And since it is usually important to use unbiased samples to obtain credible statistical results, we examine the properties that such a market should have and suggest a mechanism that compensates those individuals that participate according to their risk attitudes.  Equally important, we show that this mechanism also benefits buyers, as they pay less for the data than they would if they compensated all individuals with the same maximum fee that the most concerned ones expect.
\end{abstract}

\newpage

\section{Introduction}

Not a single week goes by without a discussion in the media and the blogosphere about data privacy in large companies and  governments, a sign that the handling of privacy is becoming a center of attention for the global society of the web.  To the usual clamor of abuses by data aggregators, marketing companies and other institutions, there is now an increased focus on the fact that, while third parties\endnote{Such as Acxiom and Quantcast} significantly profit from buying and selling consumers' data, the individuals whose data is being traded do not get paid at all. The upcoming IPO of Facebook has rekindled the discussion on whether users should benefit from the data they generate~\citep{fb_cut, fb_share}.\endnote{We note that in the case of Facebook and other online services, users do obtain a benefit in return for their data, because they get free access to these services.} And more generally, there is in principle no reason why third parties should not pay the individuals for the use of their data~\citep{new_currency, fb_aggregators}.

In a number of settings, it is of the utmost importance for the buyer of the data to obtain an {\em unbiased} sample of individuals with certain characteristics; otherwise the results of his study may not be credible.  This is usually the case for social, educational, and biomedical studies \citep[e.g.,][]{medical}.  
For instance, a pharmaceutical company might be interested in obtaining an unbiased sample of a certain demographic that have a certain disease and use a given drug.

In order to make feasible the payment to individuals for the use of their data, we design a market that consists of an intermediary (the market-maker) who facilitates the interactions between those interested in buying access to unbiased samples of certain data (the buyers) and the individuals to whom the data correspond (the sellers).
Such a market is consistent with today's reality,  whereby it is easier to obtain information from a third party than from the individual himself~\citep{clouds}.

Early suggestions for markets for private data did not specify how to set appropriate prices~\citep{secrets, laudon, coase}. More recently, these markets  were studied in the context of differential privacy~\citep{differential_privacy, differential_stratis} under the assumption that the buyer is only interested in obtaining a statistic that is based on an unbiased sample, rather than the unbiased sample itself.\endnote{We note that it is straightforward to extend our approach to settings where each buyer submits a query to the market-maker and the market-maker returns an estimate based on an unbiased sample.}

Differential privacy mechanisms are inherently randomized in the sense that the buyer does not observe the true statistic but a noisy version of it.  In this setting, potential sellers are asked to report their privacy valuations; this is, however, a major drawback from a practical point of view because it is very difficult for  individuals to know their own valuation \citep{worth}--- especially in that context.  We avoid this issue by asking sellers to choose between different pricing schemes --- instead of explicitly stating their privacy valuation.
%
Another drawback of the differentiable privacy approach is that in order to achieve a reasonably accurate estimate, the buyer needs to use data from the majority of individuals in the subset of interest, which renders this mechanism very expensive for the buyer.\endnote{For instance, \citet{differential_privacy} consider the problem of getting an unbiased estimate of the sum of $n$ bits (each of these bits corresponds to a seller).  In order to have that $Pr[|\text{estimated sum} - \text{true sum}| \geq 0.08 n] \leq 1/3$, the buyer needs to pay 95 percent of all sellers.}  Our approach avoids this problem by using small unbiased subsets of the data to compute statistics about them.

In another approach, \citet{for_sale} consider a setting where buyers pay for access to the raw data instead of just an estimate for a statistic.  They correctly argue that this is more realistic because buyers often run specialized algorithms on the data that they may not be willing to share with the market-maker.
However, \citet{for_sale} do not consider the problem of obtaining unbiased samples, which is the focus of this paper.

We consider the realistic scenario where the data corresponding to each individual is located in some database (e.g., in a hospital or a Health Information Exchange) but the buyer does not have the right to access this information.  We will assume that the individuals whose data the buyer is interested in would allow him to access this information if appropriately compensated, which elicits the question: how much should the buyer pay an individual for using his data?

As is well-known, different individuals  have different privacy valuations~\citep[see, e.g.,][]{worth}.  For instance, some individuals may not be concerned about privacy and would allow the buyer to acess their data in exchange for a few cents, whereas others may only consent if paid at least \$10.
Since all individuals prefer to be paid more,  even those  unconcerned about their privacy may pretend that they are if they expect that this will result in getting higher payments.

On the other hand, the buyer is interested in obtaining an unbiased sample without having to pay too much for it. Confronted with this problem, he may be tempted to buy access to the data with the smallest announced privacy valuations along the lines of a reverse auction. But this will not give him an unbiased sample because the value of an attribute is often correlated with its corresponding privacy valuation~\citep{weights}.
%
The requirement of an unbiased sample implies the selection of individuals with the same probability (independently of how much they value privacy),  which can significantly complicate the pricing problem.

In order to solve these problems, we describe the properties that a market for unbiased samples of private data should have, and suggest a mechanism that compensates those individuals that participate according to their own privacy and risk attitudes. Equally important, we show that this mechanism also brings benefits to buyers of the data as they end up paying for the data less than they would if they compensated all individuals with the same maximum fee that the most concerned ones expect.  Our approach is based on the fact that a significant fraction of individuals exhibit risk-averse behavior \citep{risk_aversion}.

The market-maker first collects information from the sellers with respect to what pricing schemes they prefer for allowing buyers to access their data.  When a request from a potential buyer arrives, the market-maker constructs a pricing menu for authorized, unbiased data sets that, upon purchase, the buyer can use for whatever legitimate purposes.

\section{The Market}

The market-maker facilitates the interactions between buyers and sellers according to the following sequence:

\begin{enumerate}
\item [(A)] The market-maker asks each seller a series of questions.  For each question, the seller may select between different pricing schemes that allow a future buyer to access his data.  There is always the choice to opt out.  To minimize bias, the market-maker will offer prices that are high enough so that only a negligible fraction of sellers opt out. 
\item [(B)] Every time a potential buyer provides a request specifying what type of individuals he is interested in:
\begin{enumerate}
\item [(1)]  The market-maker constructs a pricing menu for this request, which consists of the total price that the buyer has to pay in expectation to get access to an unbiased sample of $K$ individuals for various values of $K$, e.g., it may cost \$100 for $K = 100$ and \$250 for $K = 200$.
\item [(2)] The market-maker gives the pricing menu to the buyer, who then chooses the value of $K$ that he wants to buy, thus trading off accuracy and cost: a larger $K$ will provide more accurate results, but at the same time costs more.
\item [(3)] Having selected a value of $K$, the buyer pays the market-maker the corresponding price and the market-maker returns an unbiased random sample of size $K$.
\item [(4)] The market-maker uses the buyer's payment to appropriately compensate the sellers according to what they selected in (A), while keeping a cut for herself.
\end{enumerate}
\end{enumerate}

Observe that multiple buyers may use the data of the same individual.  In that case, the individual is paid each time that a buyer pays for access to the individual's data.

It is worth pointing out that to get an accurate estimate of some statistic (e.g., some average) the buyer does not need to use the entire subset that he is interested in; a sample of appropriate size will usually suffice.   In particular, let $N$ be the number of individuals who satisfy the requirements of the buyer's request and $K$ the size of an unbiased sample.  If $N$ is large, it is usually possible to obtain a good estimate even if $K \ll N$, because of the Law of Large Numbers.  

In what follows, the goal of the market-maker is to select the lowest price at which the buyer can get an unbiased sample.  In this way, the market-maker maximizes the amount of data that is traded.  Observe that if the market-maker set a high price, a seller would be better off if his data was actually sold; however, a buyer would be less likely to buy.  Thus, a high price could lead to market failure.


In the following sections, we illustrate how our approach works in a setting with two types of individuals in the population of sellers: one values privacy a lot, whereas the other does not.  We refer to individuals of the former type as high-cost and to the individuals of the latter type as low-cost.  We further assume that a low-cost individual would sell access to his data if paid any positive amount.  Experimental studies of privacy suggest that this is a good approximation of reality, as privacy valuations are clustered around two extreme values, one of which is around zero~\citep{worth}.\endnote{It is straightforward to generalize our results to the case that the privacy valuations of individuals are drawn from a general distribution as long as the support is bounded.}

\section{The Advantages of Risk Aversion}
\label{sec:aversion}

Suppose that the market-maker knows that a high-cost individual would be willing to sell access to his data if he is paid at least \$10, whereas a low-cost individual would sell access to his data if paid any positive amount.  If the market-maker knew which individuals were low-cost in advance and she had the ability to price discriminate, she would pay \$10 to high-cost individuals and a very small positive amount to low-cost individuals in exchange for selling access to their data.  But unfortunately for the market-maker and the buyers, she does not have this information.

One expensive solution is to pay \$10 to every individual whose data a buyer purchases access to.  This would certainly make low-cost individuals happy, but not the buyer.  Furthermore, a price of \$10 per data point may be too high for the buyer, resulting in a no-trade outcome.
On the other hand, if the price was strictly smaller than \$10, high-cost individuals would not participate and the resulting estimate would be biased.

The task is to find a means by which the buyer pays strictly less than \$10 on average per data point while getting an unbiased estimate.
This is possible if some individuals are risk-averse, that is, they prefer a less risky lottery to a more risky one with the same expected payment.  Experimental studies have shown that people exhibit risk-averse behavior even for small payoffs, e.g., payments equal to a couple of dollars~\citep{risk_aversion}.

Suppose that the market-maker asks each individual to select between the following two options:
\begin{itemize}
\item Option A: With probability $0.2$, a buyer will get access to your data and you will receive a payment of \$10.  Otherwise, you'll receive no payment.
\item Option B: With probability $0.2$, a buyer will get access to your data.  You'll receive a payment of \$1 irrespectively of whether or not a buyer gets access to your data.
\end{itemize}
Note that given these two options, no seller is better off opting out.
Furthermore, observe that when an individual selects Option B, he is paid even when his data is not used.

We expect that most high-cost individuals will choose Option A.  In particular, with Option A, a high-cost individual is fully compensated whenever his data is used.  On the other hand, by chosing Option B, a high-cost individual incurs a significant loss when his data is used and only receives a small payment.  We expect that this cost will generally outweigh the benefit of getting \$1 from Option B when his data is not used.

The choice of a low-cost individual will depend on his risk attitude:\endnote{The choice of a high-cost individual will also depend on his risk attitudes, but he will choose Option A for realistic risk attitudes.} he would select A if he were risk seeking or risk neutral, but would select B if he were sufficiently risk-averse.
Notice that, in expectation, the payment to someone that selects Option A is higher than the payment to someone that selects Option B, and that some individuals may select Option B because of risk aversion.

Now suppose that a buyer is interested in using a sample of individuals with certain characteristics (e.g., in terms of demographics).  Suppose that there are $N = 1000$ such individuals and the buyer wants to use a sample of size $K = 200$.
The market-maker can select a random sample of size $K$ from the population of $N$ and ask the buyer to pay each individual according to their choices.  Note that each seller is selected with probability $K/N = 0.2$. If sufficiently many sellers have selected Option B, then the expected total payment will be significantly smaller than if everyone was paid according to Option A.  For instance, if half of all individuals choose Option B, the total expected payment is $100 * 10 + 500 * 1 = 1,500$, or equivalently \$7.5 per data point.  On the other hand, if every individual is paid \$10 when his data is used, the total expected payment would be $200 * 10 = 2,000$, that is, \$10 per data point.

\section{Discovering Privacy and Risk Attitudes}
\label{sec:question}

More generally, when asking the sellers to select between potential pricing schemes, the market-maker can vary (1) the probability at which a buyer gets access to one's data, (2) the payment from Option A, and (3) the payment from Option B.  Furthermore, the market-maker can give each individual the option to opt out.  That is, an individual would be asked to select between the following three options:
\begin{itemize}
\item Option A: With probability $q$, a buyer will get access to your data and you will receive a payment of \$$x$.  Otherwise, you'll receive no payment.
\item Option B: With probability $q$, a buyer will get access to your data.  You'll receive a payment of \$$y$ irrespectively of whether or not a buyer gets access to your data.
\item Option C: No buyer gets access to your data and you receive no payment.
\end{itemize}
for various values of $q$, $x$ and $y$.  Note that, with the goal of minimizing the buyer's payment, the market-maker is only interested in values for which the payment to an individual that selects Option B is smaller than the expected payment to an individual that selects Option A, that is, $y < q \cdot x$.

Option A provides a high payment if a buyer gets access to one's data, but there is no payment otherwise.  On the other hand, Option B provides a lower strictly positive payment --- even if one's data is not used by a buyer.  Each individual will choose each time through several iterations of these three options with different prices. By aggregating the choices of all individuals, the market-maker will then be able to derive pricing menus (as described in the following section), which she can use to sell the data to buyers.

In order to select appropriate values for $x$, the market-maker can be informed by existing studies that assess how much an individual needs to be compensated to permit a decrease in privacy \citep{weights, worth, location}.  To minimize bias, the market-maker will set $x$ high enough so that only a negligible fraction of sellers opt out.

Notice that during the first iterations of market making there will be limited information as to what the ongoing price for the majority of the people is when selling access to their private data.
But as the market matures creating a pricing menu will require fewer but more accurate (in the sense of the values of $x$ and $y$) presentations to the user.  

\section{The Pricing Menu}

Let $\bar{x}$ be the minimum value of $x$ that is high enough so that only a negligible fraction of sellers has chosen Option C. In what follows we only consider the questions with $x = \bar{x}$.

When a request from a buyer arrives, the market-maker first identifies the subset of individuals that the buyer is interested in.  Suppose that this subset consists of $N$ individuals.
Then, using the choices of these individuals between Options A and B (for the questions with $x = \bar{x}$) for each value of the probability $q$, the market-maker identifies the minimum expected price for which the buyer can get access to a random unbiased sample of size $q \cdot N$.  In particular, for a fixed $q$: as $y$ increases, the number of sellers that will select Option B increases, but at the same time the expected payment to each such seller also increases.  For each probability $q$, the market-maker will choose the value $y$ that minimizes the expected amount that the buyer has to pay per data point.
By doing this for every value of $q$ for which she has the choices of sellers, the market-maker derives the pricing menu.  The pricing menu is presented to the buyer as a function of the sample size, i.e., the number of individuals in the unbiased sample.

\subsection*{Homogeneous Risk Attitudes: An Example}

To further quantify the potential gains of our approach, we consider an example where all sellers exhibit the same level of risk aversion.  We assume that the market-maker knows the distribution of privacy valuations and risk attitudes in the population of sellers but cannot price discriminate.  Note that we only make this assumption for the sake of example; in reality, the market-maker does not need to know the privacy and risk attitudes of each seller upfront, but she can discover this information as discussed in section \ref{sec:question}.

Suppose that $N = 1000$ sellers satisfy the buyer's specification.  Moreover, assume that the market-maker knows that half of these individuals are high-cost and need to be compensated at least \$10, whereas the other half are low-cost (i.e., willing to sell access to their data for any positive amount) --- but does not know who is who.  
We further assume that the risk attitudes of individuals are described by the standard utility function $u(x) = 1-e^{-x}$ and that when choosing between a triplet of options (given by the market-maker), each individual will select the option that maximizes his expected utility~\citep{MWG}.

If the market-maker has this information, then for every value of the probability $q$ (resp., the sample size $K$), she will set $x=10$ as the payment for Option A.  Moreover, she will set the payment $y$ of Option B in such a way that low-cost individuals choose it while minimizing the expected price for the buyer (the specific value of $y$ will depend on the probability $q$).  We call this the Optimal Mechanism.\endnote{In particular, at the Optimal Mechanism the payment of Option B is the minimum $y$ that satisfies $u(y) \geq q \cdot u(10) + (1-q) u(0)$ (for a fixed $q$).  This is the minimum value of $y$ for which low-cost sellers choose Option B when $x=10$.  To see this, recall that a low-cost individual incurs no cost when a buyer gets access to his data.  Thus, his utility from choosing Option B is simply $u(y)$, whereas his expected utility from choosing Option A is $q \cdot u(10) + (1-q) u(0)$ when $x = 10$.  The individual will select the option that maximizes his expected utility.}  On the other hand, we refer to the mechanism that pays any individual \$10 when his data is used as the Baseline.

\begin{figure}
\begin{center}
\includegraphics[width=8cm]{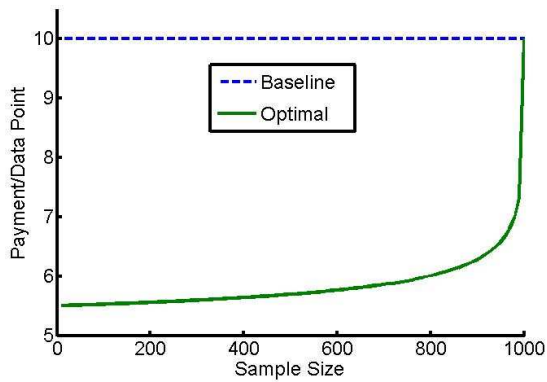}
\end{center}
\caption{Expected amount that the buyer pays per data point when sellers' risk attitudes are given by the standard utility function $u(x) = 1-e^{-x}$.}
\label{fig:comp}
\end{figure}

Figure \ref{fig:comp} shows the expected amount that the buyer pays per data point as a function of the total number of data points acquired.
We observe that, apart from very large values of $K$, the Optimal Mechanism is significantly better than the Baseline Mechanism for the buyer.  The expected payment per data point is equal to \$10 for the Baseline Mechanism, whereas for the Optimal Mechanism it may be as low as \$5.5 (for small values of $K$).  Note that if the market-maker knew the private costs of all individuals and could price discriminate, the expected payment per data point would be \$5.  We conclude that by taking advantage of sellers' risk aversion, the market-maker can significantly decrease the prices.

In this example, we have assumed that all sellers are risk-averse.  In reality, some sellers may be risk seeking, and as a result the difference between the Baseline and the Optimal Mechanism will be smaller.  But it is usually still possible to significantly decrease the price for the buyer if enough low-cost sellers are risk-averse.

\section{Bundling Across Requests}

In section \ref{sec:question}, we discussed how, by asking appropriate questions, the market-maker can discover what pricing scheme each seller prefers for selling access to his data to a buyer.  More generally, multiple buyers may pay to access the data of a single seller.  The market-maker can facilitate such trade by varying the wording in Options A and B to reflect that multiple sales may occur.  For instance, the market-maker could ask a seller to select among the following options:

\begin{itemize}
\item Option A: With probability $q$ your data will be selected.  If selected, each time a request comes for your data, the buyer that submitted it will be allowed to access your data and you will receive a payment of \$$x$.  Otherwise, you'll receive no payment and no buyer will get access to your data.
\item Option B: With probability $q$ your data will be selected.  If selected, each time a request comes for your data, the buyer that submitted it will be allowed to access your data. You'll receive a payment of \$$y$ irrespectively of whether your data is selected.
\item Option C: No buyer gets access to your data and you receive no payment.
\end{itemize}

Note that a risk-averse individual is more likely to select Option B (compared to Option A) when he expects that a larger number of buyers will be interested in his data.  As a result, the market-maker will be able to further reduce the expected payment of a single buyer if there is a large number of buyers.

Even though this approach is based on the assumption that multiple buyers will be interested in the same subset of sellers, it is possible to generalize for the case that buyers are interested in different subsets which overlap.  Then, the market-maker faces a tradeoff between (1) taking advantage of risk aversion as much as possible to minimize each buyer's expected payment and (2) asking sellers simple questions (in terms of the pricing schemes they prefer) that accurately reflect the algorithm that the market-maker uses to assign sellers to buyers.

Another related issue is whether a seller should have a say on what his data can be used for.  The market-maker can provide information about the range of applications that buyers may be interested in (e.g., biomedical, educational, and social studies).  Then, by opting in, a seller accepts that his data can be used for any of these applications by any buyer.  Alternatively, there could be multiple menus that dictate different usage policies; this approach would provide more transparency but would result in a more complex experience for the sellers.

\section{Conclusions}

In this paper we proposed a realistic and feasible market for unbiased samples of private data that compensates those individuals that opt to participate according to their own privacy and risk attitudes. This is in contrast to other market approaches to privacy that would result in the acquisition of either cheap and biased data or unbiased data sets that are large and costly. 
Our approach can be used for applications that run the gamut from biomedical to social and educational applications. Equally important, we take into account the fact that in real life a significant fraction of individuals exhibit risk-averse behavior, and we use this fact to extract a pricing structure that reflects risk attitudes and benefits the buyer of the data.

The implementation of this market would go a long way to satisfy the growing chorus of people that complain of not receiving any payment even though their data is often bought and sold by third parties at a profit.

In spite of the ease of implementation, it might take a while until individuals discover and profit from this mechanism.  But  as the demand for data increases from the buyer side of the market, we anticipate that the supply will also increase as more people realize that they can make money from selling access to their data.  Notice that in addition to potential costs because of privacy loss, an individual also has to incur the ``cost'' of answering a number of questions comparing different pricing schemes. Nevertheless it might be worthwhile to respond to these questions if people expect that their choices will be used to appropriately compensate them for a large number of sales.





\section*{Acknowledgements}

The authors gratefully acknowledge helpful conversations with Dave Levin.

\theendnotes
\bibliographystyle{abbrvnat}

\begin{thebibliography}{17}
\providecommand{\natexlab}[1]{#1}
\providecommand{\url}[1]{\texttt{#1}}
\expandafter\ifx\csname urlstyle\endcsname\relax
  \providecommand{\doi}[1]{doi: #1}\else
  \providecommand{\doi}{doi: \begingroup \urlstyle{rm}\Url}\fi

\bibitem[Acquisti et~al.(2009)Acquisti, John, and Loewenstein]{worth}
A.~Acquisti, L.~John, and G.~Loewenstein.
\newblock What is privacy worth?
\newblock In \emph{Twenty First Workshop on Information Systems and Economics
  (WISE)}, 2009.

\bibitem[Adar and Huberman(2001)]{secrets}
E.~Adar and B.~A. Huberman.
\newblock A market for secrets.
\newblock \emph{First Monday}, 6\penalty0 (8), 2001.

\bibitem[Andrews(2012)]{fb_aggregators}
L.~Andrews.
\newblock Facebook is using you.
\newblock \emph{The New York Times}, February 2012.

\bibitem[Bilton(2012)]{fb_cut}
N.~Bilton.
\newblock Disruptions: {Facebook} users ask, '{Where's} our cut?'.
\newblock \emph{The New York Times}, February 2012.

\bibitem[Cvrcek et~al.(2006)Cvrcek, Kumpost, Matyas, and Danezis]{location}
D.~Cvrcek, M.~Kumpost, V.~Matyas, and G.~Danezis.
\newblock A study on the value of location privacy.
\newblock In \emph{Proceedings of Workshop on Privacy in the Electronic Society
  (WPES '06)}, pages 109--118, 2006.

\bibitem[Dandekar et~al.(2011)Dandekar, Fawaz, and
  Ioannidis]{differential_stratis}
P.~Dandekar, N.~Fawaz, and S.~Ioannidis.
\newblock Privacy auctions for inner product disclosures.
\newblock \emph{arXiv}, 1111.2885, 2011.

\bibitem[Gellman(2009)]{clouds}
R.~Gellman.
\newblock Privacy in the clouds: Risks to privacy and confidentiality from
  cloud computing.
\newblock \emph{World Privacy Forum}, 2009.

\bibitem[Ghosh and Roth(2011)]{differential_privacy}
A.~Ghosh and A.~Roth.
\newblock Selling privacy at auction.
\newblock In \emph{ACM Conference on Electronic Commerce}, pages 199--208,
  2011.

\bibitem[Holt and Laury(2002)]{risk_aversion}
C.~A. Holt and S.~K. Laury.
\newblock Risk aversion and incentive effects.
\newblock \emph{American Economic Review}, 92:\penalty0 1644--1655, 2002.

\bibitem[Huberman et~al.(2005)Huberman, Adar, and Fine]{weights}
B.~A. Huberman, E.~Adar, and L.~R. Fine.
\newblock Valuating privacy.
\newblock \emph{Security Privacy, IEEE}, 3\penalty0 (5):\penalty0 22 -- 25,
  2005.

\bibitem[Khan et~al.(1996)Khan, Daya, and Jadad]{medical}
K.~S. Khan, S.~Daya, and A.~R. Jadad.
\newblock The importance of quality of primary studies in producing unbiased
  systematic reviews.
\newblock \emph{Arch Intern Med}, 156\penalty0 (6):\penalty0 661--666, 1996.

\bibitem[Laudon(1996)]{laudon}
K.~C. Laudon.
\newblock Markets and privacy.
\newblock \emph{Communications of the ACM}, 39, 1996.

\bibitem[Mas-Colell et~al.(1995)Mas-Colell, Whinston, and Green]{MWG}
A.~Mas-Colell, M.~D. Whinston, and J.~R. Green.
\newblock \emph{Microeconomic Theory}.
\newblock Oxford University Press, 1995.

\bibitem[Riederer et~al.(2011)Riederer, Erramilli, Chaintreau, Rodriguez, and
  Krishnamurthy]{for_sale}
C.~Riederer, V.~Erramilli, A.~Chaintreau, P.~Rodriguez, and B.~Krishnamurthy.
\newblock For sale : Your data by : You.
\newblock In \emph{HotNets}, 2011.

\bibitem[Sholtz(2001)]{coase}
P.~Sholtz.
\newblock Transaction costs and the social cost of online privacy.
\newblock \emph{First Monday}, 6\penalty0 (5), 2001.

\bibitem[Streitfeld(2012)]{fb_share}
D.~Streitfeld.
\newblock Will the {Facebook} masses demand their shares?
\newblock \emph{The New York Times}, February 2012.

\bibitem[Zax(2011)]{new_currency}
D.~Zax.
\newblock Is personal data the new currency?
\newblock \emph{Technology Review}, November 2011.

\end{thebibliography}

\end{document}